%% file: FePS3_arxiv_Haines_V2_2.tex
\documentclass[reprint,aps,prl,amsmath,amssymb]{revtex4-1}
\usepackage[T2A,LGR,T1]{fontenc}
\usepackage[latin9]{inputenc}
\usepackage{array}
\usepackage{textcomp}
\usepackage{graphicx}
\usepackage{gensymb}
\usepackage{array,booktabs}

\usepackage{dcolumn}
\usepackage{bm}

\DeclareRobustCommand{\greektext}{%
  \fontencoding{LGR}\selectfont\def\encodingdefault{LGR}}
\DeclareRobustCommand{\textgreek}[1]{\leavevmode{\greektext #1}}
\ProvideTextCommand{\~}{LGR}[1]{\char126#1}

\DeclareRobustCommand{\cyrtext}{%
  \fontencoding{T2A}\selectfont\def\encodingdefault{T2A}}
\DeclareRobustCommand{\textcyr}[1]{\leavevmode{\cyrtext #1}}

\providecommand{\tabularnewline}{\\}

\def \feps {FePS$_3$}
\newcommand{\beginsupplement}{%
        \setcounter{table}{0}
        \renewcommand{\thetable}{S\arabic{table}}%
        \setcounter{figure}{0}
        \renewcommand{\thefigure}{S\arabic{figure}}%
     }

\begin{document}

\title{Pressure induced electronic and structural phase evolution in Van der Waals compound \feps{}}

\author{C.R.S. Haines}

\affiliation{Cavendish Laboratory, Cambridge University, J.J. Thomson Ave, Cambridge
CB3 0HE, UK}

\affiliation{Department of Earth Sciences, Cambridge University, Downing Street,
Cambridge CB2 3EQ, UK}

\author{M.J. Coak}

\affiliation{Cavendish Laboratory, Cambridge University, J.J. Thomson Ave, Cambridge
CB3 0HE, UK}

\affiliation{Center for Correlated Electron Systems, Institute for Basic Science,
Seoul 08826, Republic of Korea}

\affiliation{Department of Physics and Astronomy, Seoul National University, Seoul
08826, Republic of Korea}

\author{A. R. Wildes}

\affiliation{Institut Laue-Langevin, 71  Avenue des Martyrs, 38042 Grenoble
Cedex 9, France}
\author{G. I. Lampronti}

\affiliation{Department of Earth Sciences, Cambridge University, Downing Street,
Cambridge CB2 3EQ, UK}

\author{C. Liu}

\affiliation{Cavendish Laboratory, Cambridge University, J.J. Thomson Ave, Cambridge
CB3 0HE, UK}

\author{H. Hamidov}

\affiliation{Cavendish Laboratory, Cambridge University, J.J. Thomson Ave, Cambridge
CB3 0HE, UK}

\affiliation{Navoiy Branch of the Academy of Sciences of Uzbekistan, Galaba Avenue,
Navoiy, Uzbekistan}

\affiliation{National University of Science and Technology \textquotedblleft MISiS\textquotedblright ,
Leninsky Prospekt 4, Moscow 119049, Russia}

\author{D. Daisenberger}

\affiliation{Diamond Light Source, Chilton, Didcot OX11 0DE, UK}

\author{P. Nahai-Williamson}

\affiliation{Cavendish Laboratory, Cambridge University, J.J. Thomson Ave, Cambridge
CB3 0HE, UK}

\author{S.S. Saxena}

\affiliation{Cavendish Laboratory, Cambridge University, J.J. Thomson Ave, Cambridge
CB3 0HE, UK}

\affiliation{National University of Science and Technology \textquotedblleft MISiS\textquotedblright ,
Leninsky Prospekt 4, Moscow 119049, Russia}

\date{\today}
\begin{abstract}
Two-dimensional materials have proven to be a prolific breeding ground
of new and unstudied forms of magnetism and unusual metallic states,
particularly when tuned between their insulating and metallic phases.
In this paper we present work on a new metal to insulator transition
system \feps{} . This compound is a two-dimensional van-der-Waals
antiferromagnetic Mott insulator. Here we report the discovery of
an insulator-metal transition in \feps{}, as evidenced by x-ray diffraction
and electrical transport measurements, using high pressure as a tuning
parameter. Two structural phase transitions are observed in the x-ray
diffraction data as a function of pressure and resistivity measurements
show evidence of the onset of a metallic state at high pressures. We propose
models for the two new structures that can successfully explain the
x-ray diffraction patterns.
\end{abstract}
\maketitle

\section{Introduction}
Recent and substantial interest in FePS$_3$ stems from the fact that, like graphite, FePS3 can be delaminated to give single layers.  Its magnetic properties, and particularly the Ising-like nature of its Hamiltonian, make the compound particularly promising as it maintains its bulk-like magnetic behaviour even for a single monolayer \cite{Lee_2016}. As the $ab$-planes of the
metal ions are linked only by van-der-Waals forces, the MPX$_3$ family
form model two-dimensional antiferromagnetic systems, leading to a
number of more recent publications on the magnetic structures \cite{Lancon_2016,Wildes2015,Wildes2012,Wildes2006a,Wildes1998a}
and even attracting the moniker `magnetic graphene' \cite{Park2016}.

The family of compounds here designated MPX$_3$, where M represents
a transition metal such as Fe, Ni or Mn and X = S or Se, are easily
synthesised and have been widely categorized, historically due to
interest in lithium intercalation between crystal planes to serve
as a battery material \cite{Brec1979,Rule2007,Ouvrard1985,Kurosawa1983,Klingen1968,Ichimura1991,Joy1992,LeFlem1982,Rouxel1983,Kurita1989a,Kurita1989}.
A comprehensive review is given by Grasso and Silipigni \cite{Grasso2002}.
These materials all have very similar crystal structures and interactions,
but subtly differing magnetic properties across the metal elements
due to the richness of the exchange couplings present. Crystallographic
data from x-ray diffraction studies \cite{Klingen1968,Ouvrard1985}
give a monoclinic unit cell with space group of C2/m and a honeycomb
arrangement of the magnetic metal ions.

The MPX$_3$ compounds are all p-type semiconductors \cite{Grasso2002}
with extremely high room-temperature resistivities (reflecting the
high purity of samples that can be produced) and band gaps in excess
of 1~eV as determined by optical measurements \cite{Grasso1986}. A
series of calculations of the band structure have been carried out
\cite{Whangbo1985,Kurita1989a,Zhukov1996,Piacentini1982} but do
not successfully reproduce the insulating state of these materials,
predicting half-filled metallic bands. This leads to the conclusions
that MPX$_3$ are Mott insulators, and as such could be driven to
an insulator-metal or Mott transition by applying pressure to tune
the band structure. \feps{} was chosen to investigate as it has the
lowest resistivity and band gap of the easily synthesized compounds
and was therefore assumed to require the least pressure to metallize.
Initial evidence of metallic behaviour was first presented at the SCES2014 conference in 2014
by Haines et. al. \cite{Haines2014} and then again independently
by Tsurubayashi et. al. in 2017 \cite{Tsurubayashi2017}. The current letter is the first published report showing the metal-insulator transition in FePS$_3$.

Substantial interest lies in tuning Mott insulators towards their
metallization transition. The simple Hubbard model traditionally
used to describe such systems only yields solutions in the limiting
metallic and insulating cases. Tuning the parameters of the system
to an intermediate state accesses physics that is not yet fully understood.
Additionally, many unconventional superconducting materials are low-dimensional
and lie in close proximity to antiferromagnetic Mott insulator phases
in their phase diagrams and theoretical calculations \cite{Monthoux2002}
suggest that these states have a strengthening influence on the formation
of superconductivity. Tunable (for instance through pressure)
antiferromagnetic two-dimensional Mott insulators then provide a rich
and clean environment to probe the core mechanisms of several unsolved
problems in condensed matter physics. 

\section{Methods}

High pressure XRD patterns were collected at room temperature at the
Diamond Light Source on the I15 beamline. \feps{} is difficult to
grind into an isotropic powder. This is due to the fact that when
put under strain the crystals tend to shear parallel to the $ab$-planes
leading to small platelets and therefore strong preferred orientation.
This was mitigated by grinding in liquid nitrogen. The diamond
anvil cell (DAC) was filled with powdered sample. No pressure transmitting medium was used. X-rays with an energy of 29.2~keV
(\textgreek{l} = 0.4246 \AA) were used to collect the diffraction patterns.
This is sufficiently high energy to pass through the diamond anvils.
A MAR345 2D detector was used to record the diffraction pattern with
exposure times of between 15 s and 45 s. Pressure in the DAC was determined
by measuring the shift in the fluorescence wavelength of ruby~ \cite{Mao1986}
spheres that were placed inside the high pressure region. The data
were initially processed using Dawn \cite{Filik} (with a LaB$_6$
calibration) and the subsequent Rietveld refinements were calculated
using TOPAS-Academic \cite{Topas}. 

Measurements of resistivity were performed using a Keithley 2410 Source
Meter with a fixed supplied current of 0.01~$\mu$A and verified
on a Keithley Electrometer at 40~V. As the current-voltage response
of this sample is not simply Ohmic, the same excitation was used throughout
to allow comparison of data - but limiting the measurable range of
resistances. Contact resistances to \feps{} were found to be significant
if precautions were not taken - on order of M$\Omega$ for simple
silver epoxy electrical contacts, and with significant voltage dependence.
To avoid this, the ambient pressure measurements were prepared with
gold conducting pads evaporated onto the sample surface. In the Bridgman
cells used for the high pressure resistivity measurements contact
is formed by mechanically pressing Pt wires into the sample, which
was seen to give good contact.

Measurements were performed in several Bridgman anvil cells \cite{Wittig1966}
with both tungsten carbide and sintered diamond anvils. A steatite
powder pressure medium was used in the Bridgman cells, leading to
a uniaxial component (along the sample $c*$-axis) of the pressure
achieved, as well as pressure gradients within the sample estimated
at around 20\%. The pressure in these cells is estimated from the
load applied during pressurization. 

Temperature control was achieved through an Oxford Instruments Heliox
He-3 cryostat, an ICE Oxford 1~K pumped helium cryostat and an adiabatic
demagnetization refrigerator developed in-house. Care was taken to
vary temperatures slowly (typically a few Kelvin per hour) to allow
the large thermal mass of the pressure cells to equilibrate.

\section{Results}

\begin{figure}
\begin{centering}
\includegraphics[width=1\columnwidth]{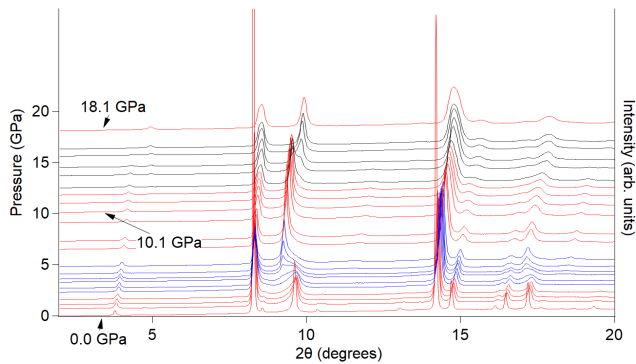}
\par\end{centering}
\caption{\label{fig:waterfall}Raw data after calibration and integration.
The data have been scaled to the low angle background (giving arbitrary
intensity, see right y-axis) and then the offset set to the pressure
at which the data were collected (shown on the left y-axis). The 0.0~GPa
data has been truncated for the two highest intensity peaks to allow
all the patterns to be plotted together. The two phase transitions
can be seen to take place over the region coloured blue (PT1) and
that coloured black (PT2). The three patterns that we identified as
being monophase are labelled with their pressures. The wavelength
of the x-rays was \textgreek{l} = 0.4246~\AA.}

\end{figure}

It was clear upon observing the evolution of the diffraction patterns,
see Fig. \ref{fig:waterfall}, that there were two phase transitions.
The first phase transition (PT1) at approximately 4~GPa and the second
(PT2) at approximately 14~GPa. The structure is known at ambient
pressure (0~GPa), however, the high pressure structures were not known.
The first task was therefore to produce physically sensible models
for the two unknown structures that could be used to fit the high
pressure diffraction patterns. In both phase transitions new diffraction
peaks develop while other peaks disappear. This allowed the identification
of monophasic diffraction patterns for all three phases: the pattern
at 0~GPa corresponds to the low pressure (ambient pressure) monoclinic
C2/m phase \cite{Klingen1968,Ouvrard1985}; the one at 10.1~GPa
to a high-pressure phase, here designated HP I; finally, the pattern
at 18.1~GPa corresponds to a second high-pressure phase, HP II. Assuming
only that the structure doesn't go through a major reconstruction
we can equate the lowest angle (highest d-spacing) peak at around
a 2$\theta$ of 3-4~\degree~ in Fig.~\ref{fig:waterfall} to the interplanar
distance. The details of the refinements can be found in the Supplementary Material.
\begin{figure}
\begin{centering}
\includegraphics[width=1\columnwidth]{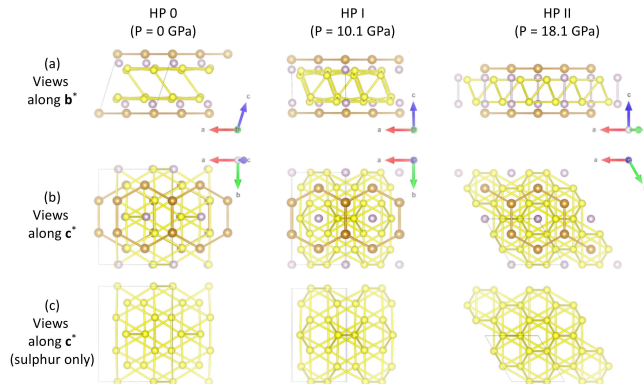}
\par\end{centering}
\caption{\label{fig:StructEvo}Schematics showing the evolution of the structure
of \feps{} with pressure. The three refined structures at their corresponding
pressures are drawn to scale. The Fe atoms are shown in brown, the
P atoms are shown in purple, and the S atoms are shown in yellow.
The views show different projections of the same number of unit cells,
hence the \textquoteright sulphur only\textquoteright{} figures show
only those sulphurs between two adjacent ab planes. Also shown are
all interatomic bonds for r <= 3.6 Angstroms. The figures were created
using the VESTA software \cite{Momma}.}
\end{figure}

Figure \ref{fig:StructEvo} shows views of the refined structures for the different phases. At ambient pressure, the sulphur atoms form two-dimensional hexagonally-coordinated
layers, and there is a close-packing coordination between the layers.
The close packing is not visible in the view along the $c*$-axis featuring
all the atoms (view (b) for HP~0 in Fig. \ref{fig:StructEvo}) because
sulphur atoms are directly below Fe and P atoms, but is visible if
only the sulphur atoms are shown (view (c) for HP~0 in Fig. \ref{fig:StructEvo}).
A shear of $\sim a/3$ along the $a$-axis, however, preserves the close-packing
between the two sulphur layers while reducing the volume of the unit
cell by putting the Fe atoms directly above one another, and likewise
for the P atoms. This is apparent in the view along $c*$, shown in
Fig.~\ref{fig:StructEvo}. The HP~I structure also appears to show some
buckling of the sulphur layers, possibly due to strain. The buckling
disappears at the higher pressure transition to HP~II, where the structure
adopts a higher symmetry. The pressure dependence of the interplanar distance is shown in Fig.~\ref{fig:StructEvo}(a).
At the second phase transition there is a dramatic
collape of the interplanar distance: from approximately 5.7~\AA~ to 4.9~\AA.
This is a collapse of nearly 15~\%. More detail can be found in the Supplementary Material.

\begin{figure}
\centering{}\includegraphics[width=1\columnwidth]{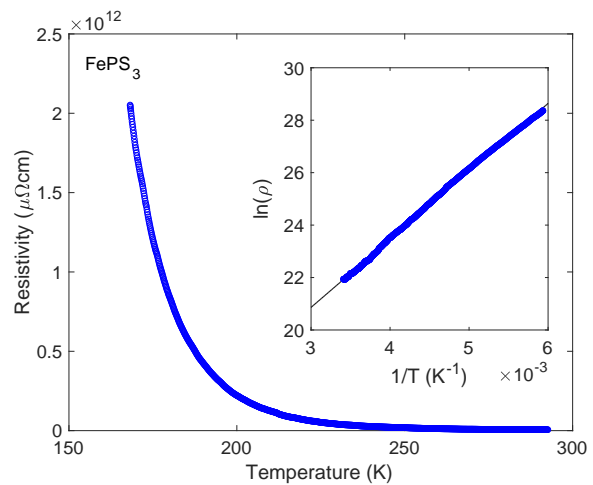}
\caption{\label{fig:FePS3_RvT_Ambient_v2}Resistivity $\rho$ of \feps{} plotted
against temperature at ambient pressure. The inset plots $\ln(\rho)$
against the reciprocal of temperature, showing good agreement with
thermally activated Arrhenius-type behaviour.}
\end{figure}

The temperature dependence of the resistivity $\rho$ of \feps{}
in the absence of any applied pressure is plotted in Fig. \ref{fig:FePS3_RvT_Ambient_v2},
from room temperature down to the temperature when the resistance
becomes too high to measure on the apparatus used. The resistivity
values found match with the orders of magnitude observed previously
\cite{Ichimura1991} and are insulating in nature. The inset shows
a fit of the resistivity to an Arrhenius-type thermally-activated
conduction process across a fixed band gap $\rho\propto e^{\frac{T_{0}}{T}}$
- plotting $\ln\rho$ against $\frac{1}{T}$ will yield a straight
line if this relation holds, which is indeed seen in the graph, with
some divergence towards the highest temperatures measured, potentially
due to carrier saturation. The extracted band gap, found from the
gradient and $T_{0},$ is then 0.452(1) eV. The overall result matches
that in previous studies \cite{Grasso1990}, but in this reference
an activation energy of 0.60(1) eV is quoted, and values up to 1.5
eV have been found from optical measurements \cite{Brec1979}. Reasons
for this mismatch are not clear, but there may be sample or orientation
dependence.

\begin{figure}[h]
\centering{}\includegraphics[width=1\columnwidth]{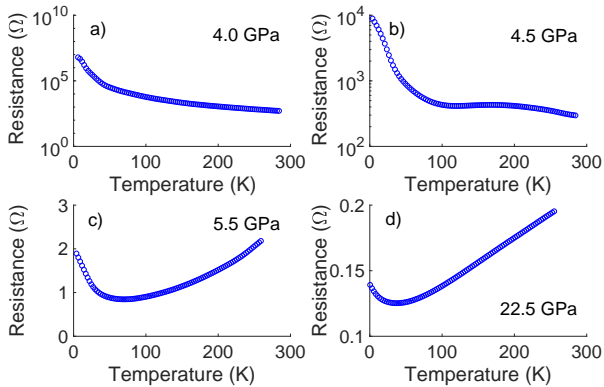}
\caption{\label{fig:FePS3_RvT_4Pressures_NahiWilliamsonUnpublished}Resistance
of \feps{} against temperature for 4 increasing pressures, estimated
as a) 4.0~GPa, b) 4.5~GPa, c) 5.5~GPa and d) 22.5~GPa. A transition
from insulating to metallic behaviour is seen as pressure is increased,
as well as an upturn in the resistivity at low temperatures in the
high pressure measurements.}
\end{figure}
\begin{figure}
\centering{}\includegraphics[width=1\columnwidth]{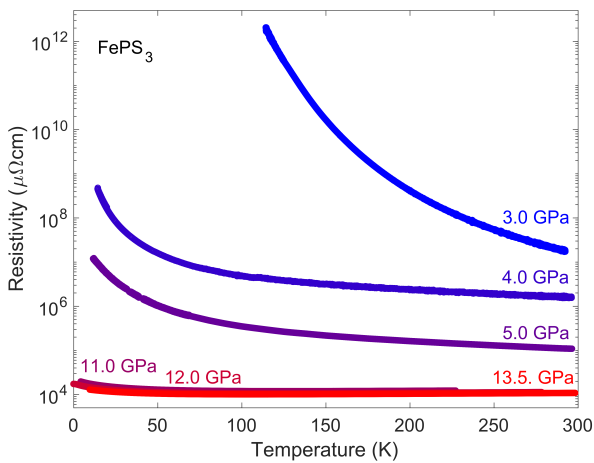}
\caption{\label{fig:FePS3_BAC_RvT-All}Resistivity of \feps{} plotted against
temperature, at pressures from an estimated 3.0 GPa (blue, topmost)
to 13.5 GPa (red) in a Bridgman anvil cell - reproducing the data
shown in Fig. \ref{fig:FePS3_RvT_4Pressures_NahiWilliamsonUnpublished}.
The resistivity is drastically suppressed with applied pressure -
note the logarithmic axis, and an upturn seen in the higher-pressure
data.}
\end{figure}

Two independent sets of measurements in the quasi-hydrostatic Bridgman
anvil cells are shown in Fig. \ref{fig:FePS3_RvT_4Pressures_NahiWilliamsonUnpublished}
and Fig. \ref{fig:FePS3_BAC_RvT-All}. In both cases the stated pressure
values are estimates based on the external load applied to the cells
in pressurization, with a typical error of up to 20\%. The lower-pressure
data (Fig. \ref{fig:FePS3_RvT_4Pressures_NahiWilliamsonUnpublished}
(a) and upper curves in Fig. \ref{fig:FePS3_BAC_RvT-All}) still show
the ambient-pressure insulating temperature dependence, but increasing
pressure drastically decreases the magnitude of the resistivity. At
pressures above an estimated 5.0~GPa the resistivity begins to decrease
with decreasing temperature - indicative of metallic behaviour. The application of pressure also leds to a decrease in the size
of the resistivity by up to 8 orders of magnitude. Fig. \ref{fig:FePS3_RvT_4Pressures_NahiWilliamsonUnpublished}
also appears to capture an intermediate state, where the insulator-metal
transition is visible as a broad peak in the resistance around 180~K,
but has moved outside of the measurable temperature range in subsequent
pressure points.

\begin{figure}
\centering{}\includegraphics[width=1\columnwidth]{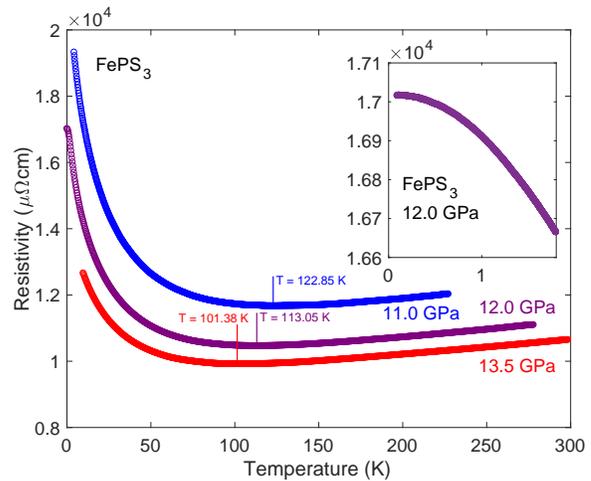}
\caption{\label{fig:FePS3_SDBAC_RvT-Inset}Detail of Fig. \ref{fig:FePS3_BAC_RvT-All}
- resistivity of \feps{} plotted against temperature, at pressures
from an estimated 11.0~GPa (blue, topmost) to 13.5~GPa (red) in
a Bridgman anvil cell. Inset shows the low-temperature data at 12.0~GPa,
which levels off or saturates at the lowest temperatures.}
\end{figure}

In the higher-pressure data taken in Bridgman cells, shown in (c)
and (d) of Fig. \ref{fig:FePS3_RvT_4Pressures_NahiWilliamsonUnpublished}
and in Fig. \ref{fig:FePS3_SDBAC_RvT-Inset}, metallic (though sub-linear
- perhaps semi-metallic) resistivity is observed at high temperatures.
At lower temperatures however, an upturn in the resistivity is seen.
This feature should not be attributed to a simple insulating state,
as the increase in resistivity as temperature decreases is slight
and far from matching the exponential dependence of an Arrhenius or
variable-range-hopping activated behaviour, or the logarithmic dependence
of the Kondo effect. The 12.0~GPa data, which were taken down to 100~mK (Fig. \ref{fig:FePS3_SDBAC_RvT-Inset},
inset), additionally show a levelling off or saturation of the resistivity
as the temperature approaches zero; this resembles the saturation effects commonly seen in
semiconductors. What can however be concluded from Fig. \ref{fig:FePS3_BAC_RvT-All}
is that as pressure is increased this upturn feature is suppressed
in size and temperature scale. \feps{} at ambient pressure orders
antiferromagnetically at 120 K, and it is tempting to speculate that
the upturn may also have some connection to the magnetic properties
of the compound at high pressure.

\section{Discussion}

The application of pressure changes the structure and electronic properties
of \feps{}. Through the analysis of powder x-ray diffraction data
under pressure we have discovered and produced models for two previously
unknown phases of \feps{}. The change from the ambient pressure structure
to HPI amounts to a shear of the unit cell by $\sim$a/3 along the
$a$-axis. A shear between the $ab$-planes with pressure seems feasible
as the planes are only weakly bound by van der Waals forces and they
slide easily over one another. Inspection of the stacking normal to
the ab planes, i.e. along the $c*$-axis, for the different structures
supports a coherent shear model. 

Under ambient conditions, the layered van-der-Waals antiferromagnet
\feps{} displays insulating behaviour, with an electronic band gap
around 0.5~eV, in agreement with previous studies. The application
of comparatively low pressures, up to 2.0~GPa, has little effect
on the resistivity but higher pressures reduce the sample's resistivity
values at a dramatic rate. In the quasi-hydrostatic Bridgman cells,
a reduction in sample resistance values from G$\Omega$ to $\Omega$
ranges is seen, and a resistance increasing with temperature from
pressures around 5.0~GPa, indicative of a metallic rather than insulating
state. Electronic structure calculations suggest \feps{} to be a
Mott or Charge-Transfer insulator, and this insulator-metal transition
observed through applied pressure strongly supports this view. Additionally,
in the metallic phase at pressures above the transition, an upturn
in the resistivity is observed at temperatures below 150~K, suppressed
with increasing pressure. The upturn could not be described by any
common forms of exponential insulator type temperature dependencies
or scattering mechanisms such as the Kondo effect. Further information
into the band structure and nature of carriers in this high-pressure
state is needed to properly understand this feature. 

We show that \feps{} undergoes an insulator-metal transition at high pressures and we are able to correlate the electronic transition with concomitant changes in the crystallographic structure.  The experiments are challenging, but the data are unambiguous and conclusive.

A strain or pressure induced transition to a metallic phase is not only relevant for the graphene community.  The transition of a layered antiferromagnetic insulator to a conductor is also found in high-temperature superconductors, which is a subject of long-standing interest.  The electronic properties of high-temperature superconductors are controlled by doping, and it is very difficult to unravel the effects that come into play in non-stoichiometric systems.  Our study suggests another avenue, as the application of pressure to a stoichiometric system provide causal relations that are far less ambiguous, and we believe that pressure studies on \feps{} may also provide insight into the origins of high-temperature superconductivity.

\section*{Acknowledgements}

\begin{acknowledgments}
This work was carried out with the support of the Diamond Light Source
and we thank Heribert Wilhelm for advice and help with the X-ray diffraction
experiments at Diamond. The authors would like to thank G.G. Lonzarich,
P.A.C. Brown, S.E. Dutton, L.J. Spalek and D. Jarvis for their help
and discussions. We would also like to acknowledge support from Jesus
College of the University of Cambridge, the Engineering and Physical
Sciences Research Council, IHT KAZATAMPROM and the CHT Uzbekistan
programme. The work was carried out with financial support from the
Ministry of Education and Science of the Russian Federation in the
framework of Increase Competitiveness Program of NUST MISiS (\textnumero{}
\textcyr{\CYRK}2-2017-024). This work was supported by IBS-R009-G1.
\end{acknowledgments}


%

\section*{Supplementary Material}

\beginsupplement
\input{supplementary}

\end{document}

%% file: supplementary.tex
\subsection{Crystallography and solving the new structures of \feps}
\begin{figure}
\begin{centering}
\includegraphics[width=1\columnwidth]{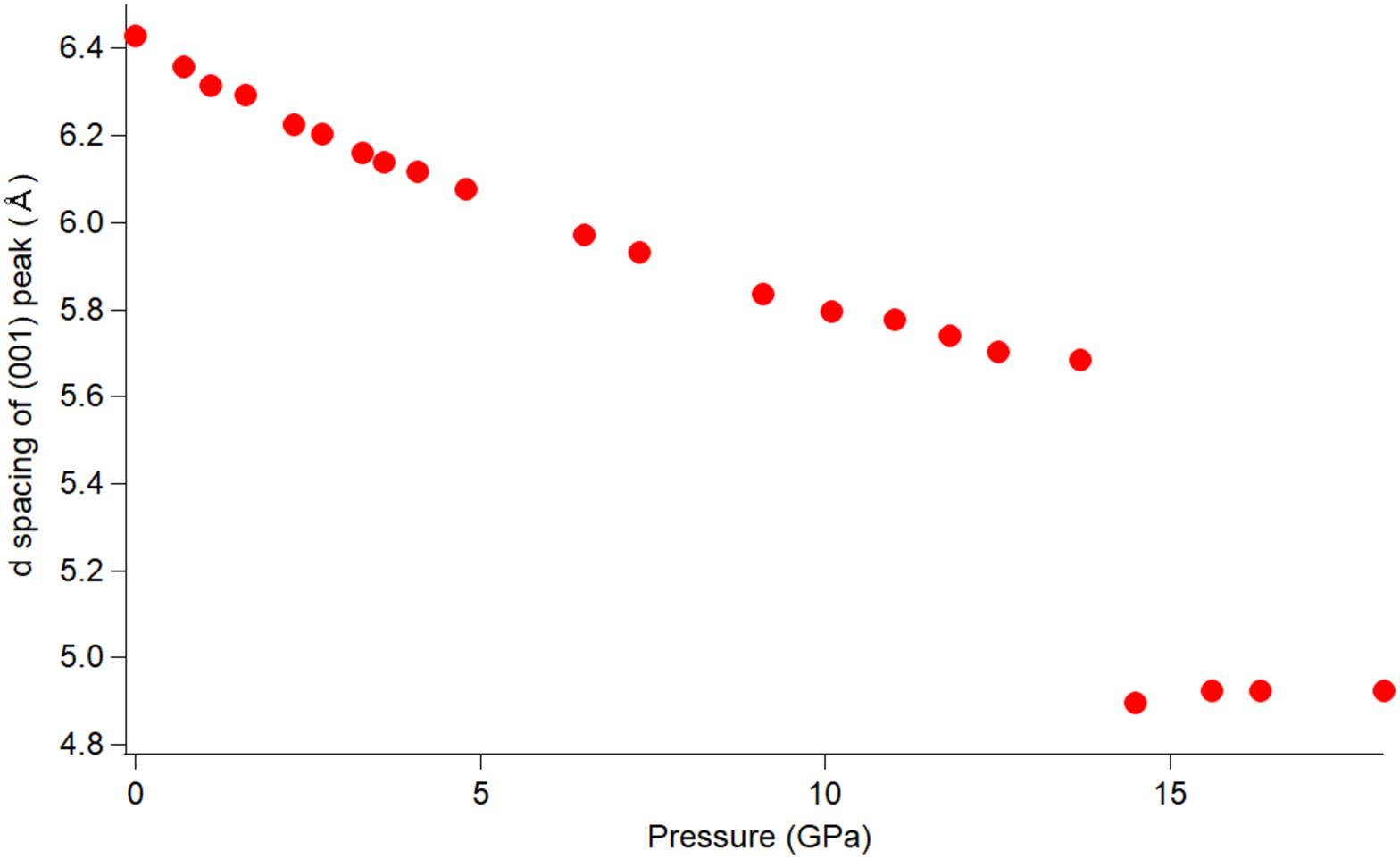}
\par\end{centering}
\caption{\label{fig:c_spacing}The interplanar distance is given by the (001)
d spacing. This distance is reduced smoothly by applied pressure until
the second phase transition at $\sim$14~GPa when it collapses dramatically.}

\end{figure}

\begin{figure}
\begin{centering}
\includegraphics[width=1\columnwidth,height=0.6\columnwidth]{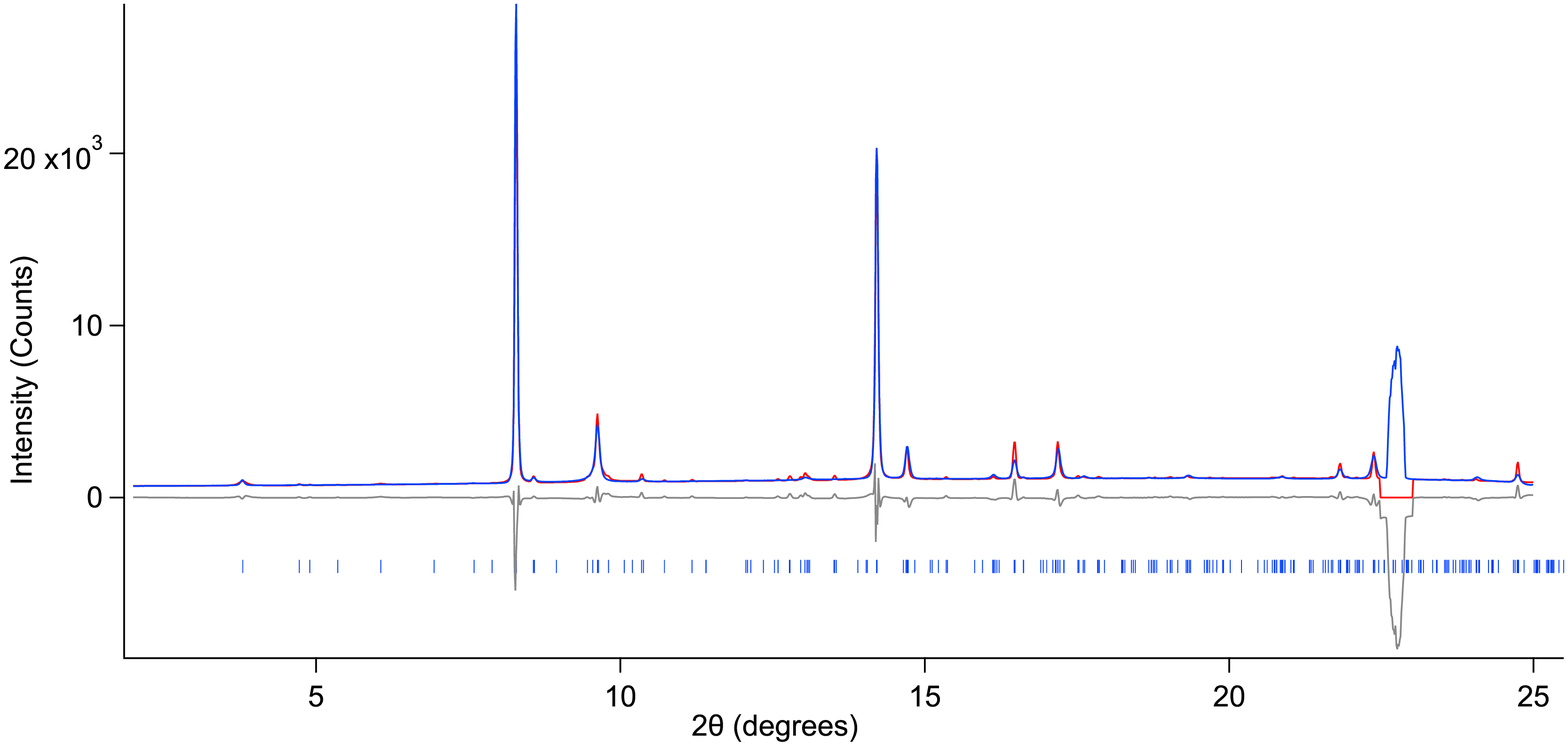}
\par\end{centering}
\caption{\label{fig:0GPaRefine}Rietveld refinement plot of 0 GPa diffraction
pattern with known C2/m structure showing experimental (blue), calculated
(red) and difference (grey) curves. Reflection positions are marked
in blue. The large peak at approximately 22.5\degree is from the diamond
anvils of the pressure cell. The region immediately around the peak
has been excluded from the fit. The wavelength of the x-rays was \textgreek{l}
= 0.4246 \AA.}
\end{figure}

The first phase transition seems to lead to fewer peaks, suggesting
higher symmetry, but as yet we still find the best match with a modified
monoclinic C2/m structure. This is closely related to the starting
structure, with the primary difference being a change in \textgreek{b}
from  $\sim$107\textdegree{} to  $\sim$90\textdegree . Additionally,
the P-P distance increases, reflecting the change in the atomic stacking
along the c{*} axis. The P-P pairs are separated by a sulphur atom
at 0 GPa, while the ab-planes have sheared at higher pressures such
that this is no longer the case and P-P pairs form a chain along c{*}.
The change in the c{*} stacking is shown in Fig.\ref{fig:StructEvo}.

\begin{figure}
\begin{centering}
\includegraphics[width=1\columnwidth,height=0.6\columnwidth]{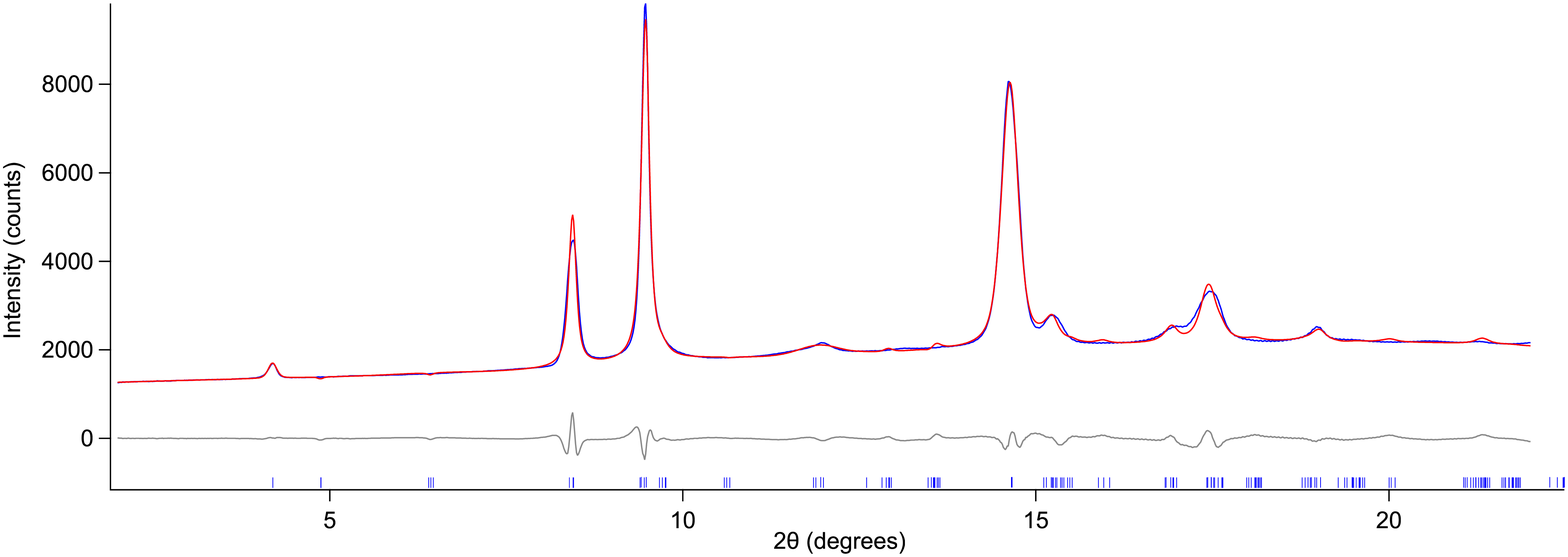}
\par\end{centering}
\caption{\label{fig:10GPaRefine}Rietveld refinement plot of 10.1 GPa diffraction
pattern with proposed C2/m structure for the 1st high pressure phase
- HPI showing experimental (blue), calculated (red) and difference
(grey) curves. Reflection positions are marked in blue. The wavelength
of the x-rays was \textgreek{l} = 0.4246 \AA.}
\end{figure}

The second phase transition then leads to the closely related P-31m
trigonal structure. The principal changes here are the increase in
symmetry and the collapse of the interlayer spacing. 

\begin{figure}
\begin{centering}
\includegraphics[width=1\columnwidth,height=0.6\columnwidth]{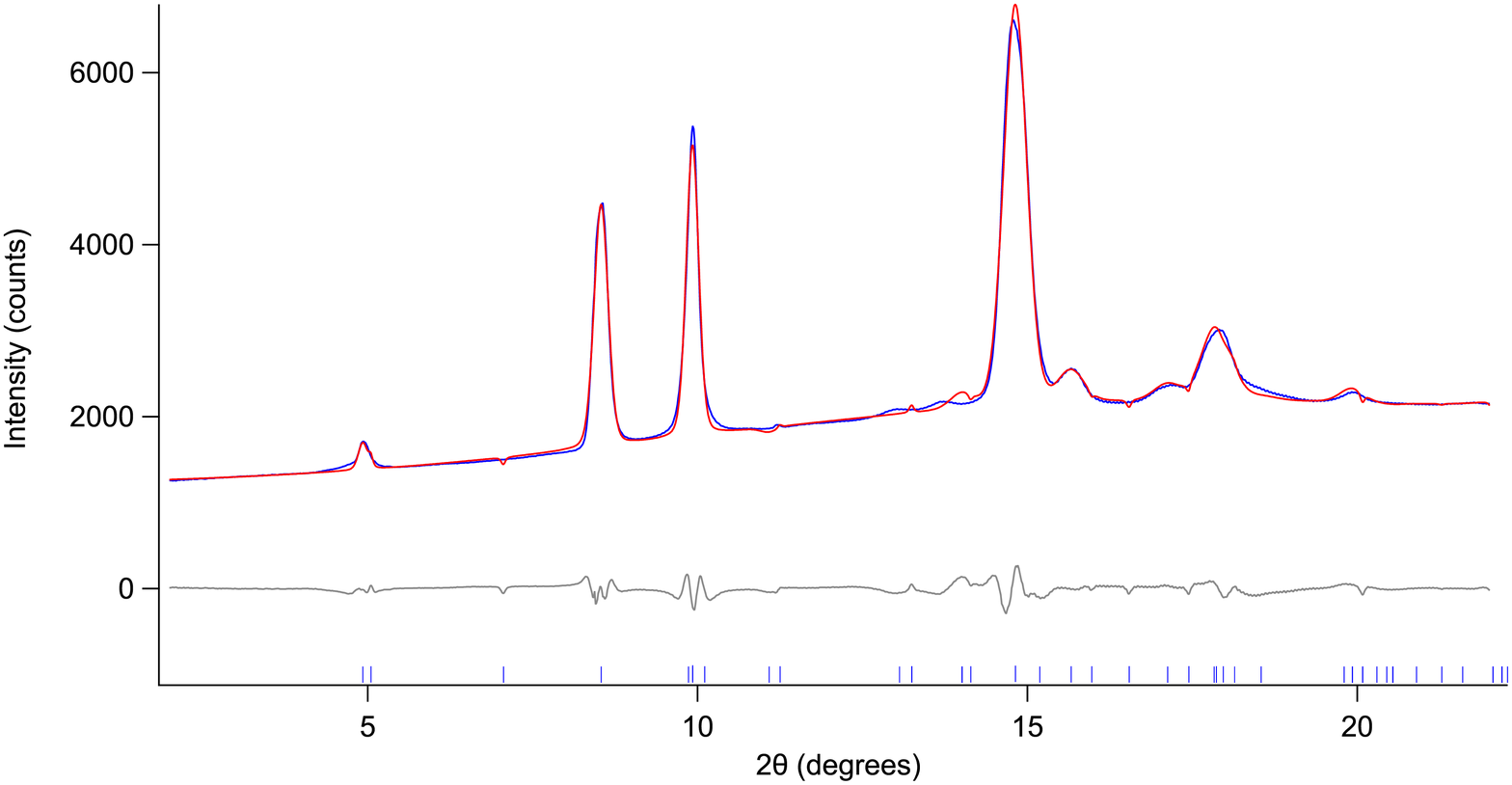}
\par\end{centering}
\caption{\label{fig:18GPaRefine}Rietveld refinement plot of 18.1 GPa diffraction
pattern with proposed P-31m structure for the 2nd high pressure phase
- HPII showing experimental (blue), calculated (red) and difference
(grey) curves. Reflection positions are marked in blue. The wavelength
of the x-rays was \textgreek{l} = 0.4246~\AA. }
\end{figure}

\begin{table*}
\begin{centering}
\begin{tabular}{>{\centering}p{0.14\textwidth}>{\centering}p{0.14\textwidth}>{\centering}p{0.14\textwidth}>{\centering}p{0.14\textwidth}>{\centering}p{0.16\textwidth}>{\centering}p{0.16\textwidth}}
\toprule 
\addlinespace[0.2cm]
\multicolumn{2}{c}{Ouvrard 1985 } & C2/m & R = 4\% & R$_{w}$= 5\% & \tabularnewline
\addlinespace[0.4cm]
a = 5.947(1)~\AA & b = 10.300(1)~\AA & c = 6.7222(8)~\AA & \textgreek{b} = 107.16(1)~\degree & V = 393.3(2)~\AA$^{3}$ & \tabularnewline
\addlinespace[0.2cm]
\midrule 
 & x & y & z & Occ & B$_{eq}$\tabularnewline
\midrule
Fe(4g) & 0 & 0.3326(1) &  0 & 1 & 1.11(1)\tabularnewline
\addlinespace[0.1cm]
P(4i) & 0.0566(3) & 0 & 0.1692(3) & 1 & 0.65(2)\tabularnewline
\addlinespace[0.1cm]
S(4i) & 0.7501(3) & 0 & 0.2470(3) & 1 & 0.88(2)\tabularnewline
\addlinespace[0.1cm]
S(8j) & 0.2488(2) & 0.1655(1) & 0.2485(2) & 1 & 0.87(1)\tabularnewline
\bottomrule
\addlinespace[0.1cm]
\end{tabular}
\par\end{centering}
\caption{\label{tab:ref_0gpa}Refinement parameters from Ouvrard\cite{Ouvrard1985}.}
\end{table*}

\begin{table*}
\begin{centering}
\begin{tabular}{>{\centering}p{0.14\textwidth}>{\centering}p{0.14\textwidth}>{\centering}p{0.14\textwidth}>{\centering}p{0.14\textwidth}>{\centering}p{0.16\textwidth}>{\centering}p{0.16\textwidth}}
\toprule 
\addlinespace[0.2cm]
\multicolumn{2}{c}{HP 0 } & C2/m & r$_{wp}$= 8.07 & r$_{exp}$= 2.92 & \textgreek{q}$^{2}$= 2.77\tabularnewline
\addlinespace[0.2cm]
a = 5.9428(9)~\AA & b = 10.299(2)~\AA & c = 6.716(2)~\AA & \textgreek{b} = 107.34(2)~\degree & V = 392.4(2)~\AA$^{3}$ & \textgreek{r} = 3.098(1)~g.cm$^{-3}$\tabularnewline
\addlinespace[0.2cm]
\midrule 
 & x & y & z & Occ & B$_{eq}$\tabularnewline
\midrule
Fe(4g) & 0 & 0.3320(8) &  0 & 1 & 3.4(2)\tabularnewline
\addlinespace[0.1cm]
P(4i) & 0.086(4) & 0 & 0.167(7) & 1 & 4.2(6)\tabularnewline
\addlinespace[0.1cm]
S(4i) & 0.760(4) & 0 & 0.286(6) & 1 & 2.99(19)\tabularnewline
\addlinespace[0.1cm]
S(8j) & 0.269(3) & 0.1745(9) & 0.247(4) & 1 & 2.99(19)\tabularnewline
\bottomrule
\addlinespace[0.1cm]
\end{tabular}
\par\end{centering}
\caption{\label{tab:ref_0gpa-hp0}Refined crystal structure parameters for
the 0 GPa (HP 0) data.}
\end{table*}

\begin{table*}
\begin{centering}
\begin{tabular}{>{\centering}p{0.14\textwidth}>{\centering}p{0.14\textwidth}>{\centering}p{0.14\textwidth}>{\centering}p{0.14\textwidth}>{\centering}p{0.16\textwidth}>{\centering}p{0.16\textwidth}}
\toprule 
\addlinespace[0.2cm]
\multicolumn{2}{c}{HP I } & C2/m & r$_{wp}$= 2.45 & r$_{exp}$= 2.14 & \textgreek{q}$^{2}$= 1.15\tabularnewline
\addlinespace[0.2cm]
a = 5.7620(12)~\AA & b = 9.988(2)~\AA & c = 5.803(5)~\AA & \textgreek{b} = 89.33(2)~\degree & V = 333.3(3)~\AA$^{3}$ & \textgreek{r} = 3.648(3)~g.cm$^{-3}$\tabularnewline
\addlinespace[0.2cm]
\midrule 
 & x & y & z & Occ & B$_{eq}$\tabularnewline
\midrule
Fe(4g) & 0 & 0.3225(13) &  0 & 1 & 1.0(10)\tabularnewline
\addlinespace[0.2cm]
P(4i) & 0 & 0 & 0.184(15) & 1 & 0.1(12)\tabularnewline
\addlinespace[0.2cm]
S(4i) & 0.638(4) & 0 & 0.259(13) & 1 & 0.1(6)\tabularnewline
\addlinespace[0.2cm]
S(8j) & 0.127(2) & 0.16239 & 0.299(8) & 1 & 0.1(6)\tabularnewline
\bottomrule
\addlinespace[0.2cm]
\end{tabular}
\par\end{centering}
\caption{\label{tab:ref_highp}Refined crystal structure parameters for the
HP I phase (10 GPa data).}
\end{table*}

\begin{table*}
\begin{centering}
\begin{tabular}{>{\centering}p{0.14\textwidth}>{\centering}p{0.14\textwidth}>{\centering}p{0.14\textwidth}>{\centering}p{0.14\textwidth}>{\centering}p{0.16\textwidth}>{\centering}p{0.16\textwidth}}
\toprule 
\addlinespace[0.2cm]
\multicolumn{2}{c}{HP II } & P-31m & r$_{wp}$= 2.06 & r$_{exp}$= 2.14 & \textgreek{q}$^{2}$= 0.95\tabularnewline
\addlinespace[0.2cm]
a = 5.699(4)~\AA &  & c = 4.818(3)~\AA &  & V = 135.54(19)~\AA$^{3}$ & \textgreek{r} = 4.484(6)~g.cm$^{-3}$\tabularnewline
\addlinespace[0.2cm]
\midrule 
 & x & y & z & Occ & B$_{eq}$\tabularnewline
\midrule
Fe(4g) & 1/3 & 2/3 &  0 & 1 & 1.0(2)\tabularnewline
\addlinespace[0.2cm]
P(4i) & 0 & 0 & -0.206(6) & 1 & 1.0(6)\tabularnewline
\addlinespace[0.2cm]
S(4i) & 0.3241(6) & 0 & -0.195(4) & 1 & 1.0(2)\tabularnewline
\bottomrule
\addlinespace[0.2cm]
\end{tabular}
\par\end{centering}
\caption{\label{tab:ref_highp2}Refined crystal structure parameters for the
HP II phase (18 GPa data).}
\end{table*}

Rietveld refinements were performed using Topas V6. A shifted Chebyshev
function with 6 parameters was used to fit the background. The Pseudo-Voigt
function used to model the peak shape and the parameters describing
the diffractometer geometry were first optimized using an LaB$_{6}$
standard. These were fixed for the structural refinements, while two
isotropic parameters were used to take into account the sample Lorentzian
contribution to peak broadening for size and microstrain respectively
unless otherwise stated. Structural parameters were refined with an
anti-bump restraint for the P-P interatomic distance. A minimum value
of 0.1 \AA$^{2}$ was set for the thermal parameters (one for each atomic
species). Goodness of fit indices, R$_{wp}$ values and structural
parameters at convergence are reported in Tables \ref{tab:ref_0gpa}
and \ref{tab:ref_highp}. Final Rietveld refinement plots are shown
in Figures \ref{fig:0GPaRefine}, \ref{fig:10GPaRefine} and \ref{fig:18GPaRefine}
respectively. The ambient \feps{} crystal structure \cite{Ouvrard1985}
was retrieved from the ICSD \cite{Allmann2007} to fit the 0 GPa
powder pattern. The March-Dollase model for preferred orientation
was applied on the (0 0 1) crystallographic plane. An anisotropic
model \cite{Stephens1999} for sample microstrain contribution was
found to improve the fit significantly. The anisotropic broadening
changes with increasing pressure (see Figure \ref{fig:0GPaRefine}
from 8 to 10\degree) and it is likely to be related to stacking faults phenomena.
We are currently working on a stacking faults model that will describe
these changes in detail. The R$_{Bragg}$ was found to be 7.35 at
convergence. A few minor peaks disappear in the diffraction patterns
when pressure increases to 10 GPa (see the 5 to 10\degree range in Fig.\ref{fig:10GPaRefine}).
Because of this apparent increase in symmetry, indexing in cubic,
hexagonal, trigonal, tetragonal and orthorhombic symmetries was extensively
attempted on the 10 GPa diffraction pattern with the indexing algorithm
DICVOL \cite{Boultif2004}, which is included in the software DASH
\cite{David2006} from the CCDC suite\cite{Groom2016}. One hexagonal
unit cell was found to be rather convincing but a Pawley fit proved
that it fails to fit the (0 0 1) peak properly (see Figure 1 in the
SI). The pattern can instead be fitted with an orthorhombic C-centered
unit cell (see Figure 1 in the SI). However, so far we have been unable
to model the structure with an orthorhombic supergroup of C2/m. Therefore,
our conclusion is that the HP I crystal structure is in fact monoclinic
with the same space group as the 0 GPa crystal structure. We believe
that the stacking faults model we are currently working on could shed
more light on the structural behaviour during this phase transition.
An 8th order spherical harmonics model for preferred orientation was
found to be necessary to fit the peak intensities in this case. The
R$_{Bragg}$ was found to be 0.47 at convergence. A Pawley fit showed
that the 18.1 GPa diffraction pattern could be fitted with the hexagonal
unit cell identified during the indexing of the 10 GPa pattern. A
manual determination of systematic extinctions excluded any reflection
condition. A structural model was built assuming that the layers would
be perpendicular to {[}0 0 1{]} with the P atoms lying on the 3-fold
axis and stack one onto the other without any shift on the a,b plane
(see Fig.\ref{fig:StructEvo}). The space group P-31m, which has no
systematic absences, was found to be compatible with this structural
model. An 8th order spherical harmonics model for preferred orientation
was used. The R$_{Bragg}$ was found to be 0.44 at convergence. 

\begin{figure}
\begin{centering}
\includegraphics[width=1\columnwidth]{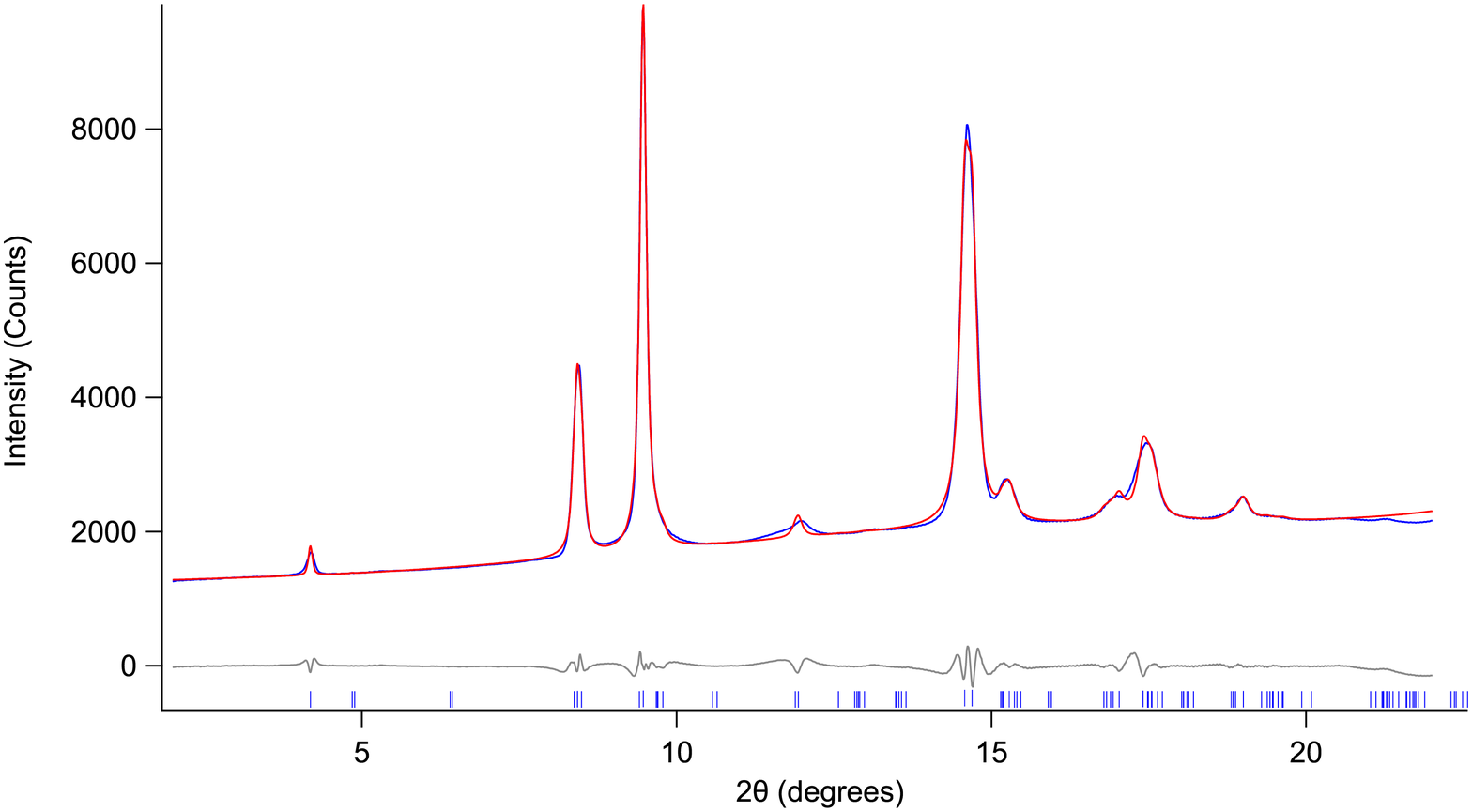}
\par\end{centering}
\begin{centering}
\includegraphics[width=1\columnwidth]{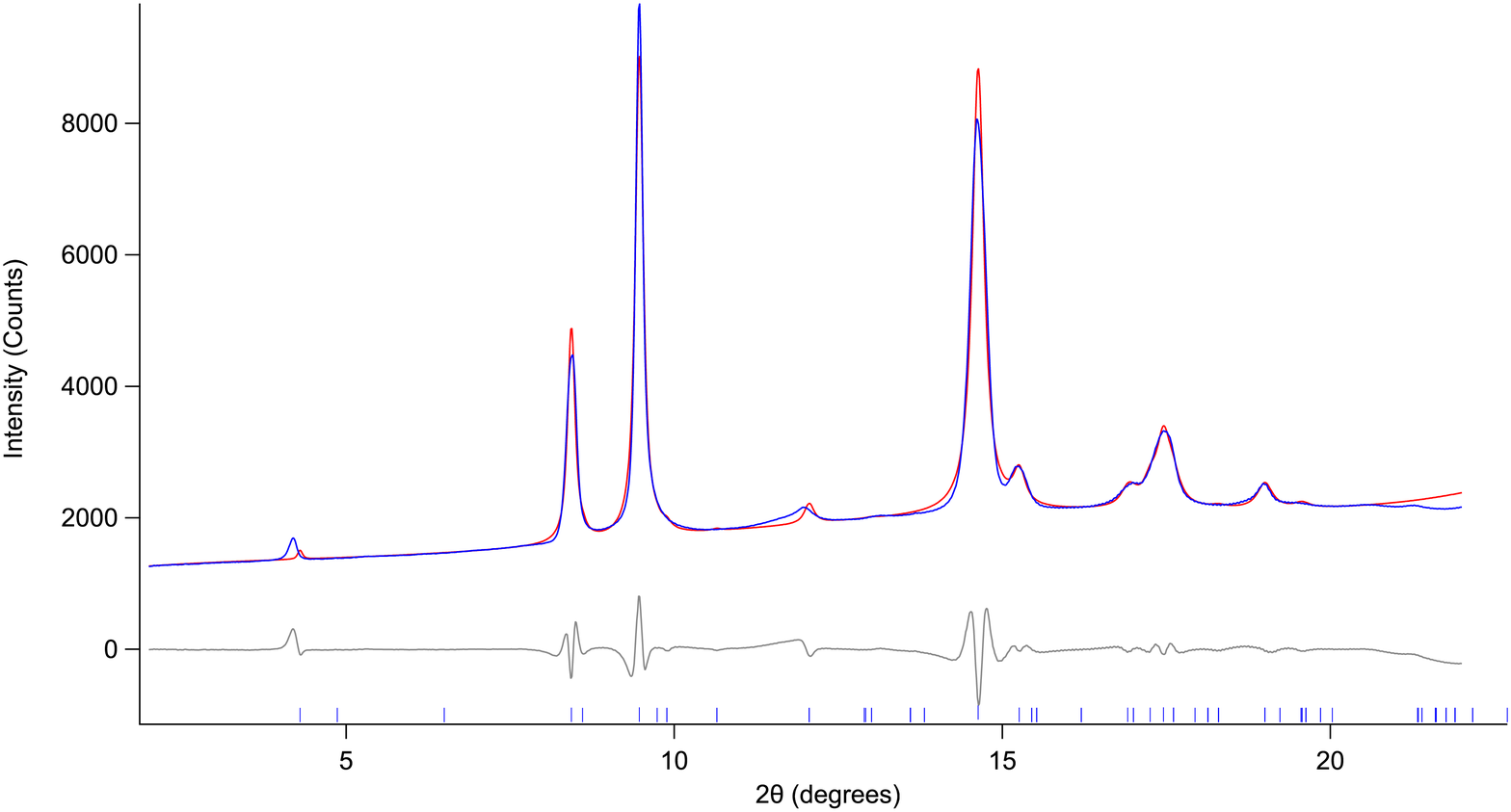}
\par\end{centering}
\caption{\label{fig:10GPa_pawleyCmmm}Pawley fit plot of 10.1 GPa diffraction
pattern with the identified potential Cmmm (upper) and P3 (lower)
symmetries for the 1st high pressure phase - HPI showing experimental
(blue), calculated (red) and difference (grey) curves. Reflection
positions are marked in blue.}

\end{figure}